\documentclass[12pt,a4paper,final]{iopart}

\usepackage{iopams}  
\expandafter\let\csname equation*\endcsname\relax
\expandafter\let\csname endequation*\endcsname\relax
\usepackage{amsmath}
\usepackage{amssymb}
\usepackage{graphicx}
\graphicspath{ {./images/} }
\usepackage{cite}
\usepackage{float}
\usepackage[breaklinks=true,colorlinks=true,linkcolor=blue,urlcolor=blue,citecolor=blue]{hyperref}

\newcommand{\be}{\begin{equation}}
\newcommand{\ee}{\end{equation}}
\newcommand{\bal}{\begin{aligned}}
\newcommand{\eal}{\end{aligned}}

\newcommand\norm[1]{\left\lVert#1\right\rVert}

\begin{document}

\title{Lorentzian Spectral Geometry with Causal Sets}

\author{Yasaman K. Yazdi$^{1,2}$, Marco Letizia$^{3,4,5}$ and  Achim Kempf$^{3,4,6}$}

\address{$^1$Theoretical  Physics  Group, Blackett  Laboratory,  Imperial  College London, 
\\SW7 2AZ, UK}
\address{$^2$Department of Physics, 4-181 CCIS, University of Alberta, Edmonton AB, \\T6G 2E1, Canada}
\address{$^3$Dept. of Applied Mathematics, Univ. of Waterloo, Waterloo, ON, N2L 3G1, Canada}
\address{$^4$Perimeter Institute for Theoretical Physics, Waterloo, ON, N2L 2Y5, Canada}
\address{$^5$MaLGa, DIBRIS, University of Genoa, Via Dodecaneso 35, 16146 Genoa, Italy}
\address{$^6$Inst. for Quantum Computing, Univ. of Waterloo, Waterloo, ON, N2L 3G1, Canada}
\ead{ykouchek@imperial.ac.uk, marco.letizia@edu.unige.it,  akempf@pitp.ca}

\begin{abstract}
We study discrete Lorentzian spectral geometry by investigating to what extent causal sets can be identified through a set of  geometric invariants such as spectra. We build on previous work where it was shown that the spectra of certain operators derived from the causal matrix possess considerable but not complete power to distinguish causal sets. We find two especially successful methods for classifying causal sets and we computationally test them for all causal sets of up to $9$ elements. One of the spectral geometric methods that we study involves holding a given causal set fixed and collecting a growing set of its geometric invariants such as spectra (including the spectra of the commutator of certain operators). The second method involves obtaining a limited set of geometric invariants for a given causal set while also collecting these geometric invariants for small `perturbations' of the causal set, a novel method that may also be useful in other areas of spectral geometry. We show that with a suitably chosen set of geometric invariants, this new method fully resolves the causal sets we considered. Concretely, we consider for this purpose perturbations of the original causal set that are formed by adding one element and a link. We discuss potential applications to the path integral in quantum gravity.
\end{abstract}

\renewcommand*\contentsname{}

\section{Introduction}

Understanding quantum gravity is one of the most challenging open questions in modern theoretical physics. Several different approaches to this problem gave rise to active areas of research. These approaches include, for example, causal set theory \cite{SuryaRev}, causal dynamical triangulations \cite{Loll:2019rdj}, loop quantum gravity \cite{rovelli_2004}, spinfoams \cite{rovelli_vidotto_2014} and asymptotic safety \cite{Niedermaier_2007}. One of the questions that each of these programs faces is how to uniquely identify the  gravitational degrees of freedom for the purpose of quantization. In making this identification, one needs to carefully exclude gauge degrees of freedom such as choices of coordinates. In continuum and discrete spacetime theories, this means that the diffeomorphism group or the permutation group need to be modded out respectively. An alternative to modding out the diffeomorphism or permutation group is to work only with invariants from the start. This raises the challenge to identify a sufficient number of geometric invariants so that they together allow one to identify all the physical spacetime degrees of freedom. 

One natural set of invariants are the eigenvalues of an operator because, regardless of which   orthonormal basis (for example different coordinate systems) the operator is expressed in, its  eigenvalues will take the same values. Spectral geometry uses this fact to characterize geometric information, for example, about a compact Riemannian manifold through the spectra of operators such as the Laplace operators on the manifold.\cite{davies,berard2006spectral,craioveanu2013old}. In this context, we are here following up on a prior paper by some of the present authors \cite{SG1}.

\subsection{The Challenge of Spectral Geometry}
The key challenge in the field of spectral geometry, and the motivation for the present work, is the observation that in many circumstances the knowledge of the spectrum of an operator such as the Laplacian on scalar functions is insufficient to determine the metric of the underlying manifold, such as that of a compact Riemannian manifold. 
It is instructive to discuss the reasons for why the spectrum of an operator such as the Laplacian may naively be expected to contain all information about the metric, and the reasons for why it nevertheless often does not. 

First, it is known that if the Green function $G(x,x')$ of a massless scalar field on a Riemannian or pseudo-Riemannian manifold is known then the metric of the manifold can always be reconstructed explicitly from the Green function, see \cite{Saravani:2015moa}. This is possible because the latter provides a measure of the covariant distance between $x$ and $x'$ and this includes a measure of the infinitesimal distances encoded by the metric. The Green function is the left inverse of the Laplacian operator, e.g., on Riemannian manifolds. This means that if the Laplacian is known as an operator acting on scalar functions over the manifold, that is to say, if the Laplacian is known in the position basis, then the metric can be calculated from it. This is done by first calculating its inverse which is the Green function, and by then calculating from it the metric. 

Naively, this suggests  that the spectrum of the Laplace operator fully encodes the metric of the manifold.
The reason is that while the Laplace operator, as a differential operator on functions, fully encodes the metric, it does so in a highly redundant way. This is because the same metric manifold can be described by infinitely many Laplace operators that differ merely by a change of coordinates on the manifold, viz., by a diffeomorphism. The explicit form that a Laplace operator takes as a differential operator on scalar functions in some coordinate system expresses not only the underlying metric but expresses in large part the choice of coordinates. For the purposes of quantum gravity it is of course only the actual metric information in the Laplacian that is of interest. 

This is where the spectrum of the Laplacian becomes of interest, because it is diffeomorphism invariant. Intuitively, the vibration spectrum of a manifold does not depend on the coordinate system used to calculate it. Mathematically, any change of coordinates, i.e., any diffeomorphism is merely a change of basis in the Hilbert space of square integrable functions on the manifold and it therefore leaves the spectrum of the Laplacian invariant. The spectrum of the Laplacian is, therefore, a description of metric degrees of freedom of the manifold that is nonredundant. It is diffeomorphism invariant, and therefore the Laplacian's eigenvalues are geometric invariants.  

However, does the spectrum of the Laplacian also provide a complete set of the geometric invariants, i.e., does it suffice to reconstruct the metric fully? One might expect that the answer is yes because the spectrum is the full set of invariants of an operator under the full unitary group of the Hilbert space. Diffeomorphisms are changes of basis in the Hilbert space and are, therefore, elements of the full unitary group. Nevertheless, as is well known, the spectrum of the Laplacian generally does not suffice to reconstruct the metric completely, see e.g., \cite{Kempf_2010,datchev2011inverse,kempf2018quantum}. How can this be? 

The reason is that the diffeomorphism group, while contained in the unitary group of the Hilbert space, is much smaller than the unitary group. For example, Fourier transforms are unitary changes of basis but they are not diffeomorphisms. This means that when we give up the Laplacian in a coordinate system and retain only its spectrum, then we are in effect extracting only those invariants of the diffeomorphism group that are also invariants of the larger unitary group. 

If we were able to extract from the Laplacian in a coordinate system the set of all quantities that are only invariant under the diffeomorphism group (and not necessarily the full unitary group) then they would  suffice to recover the metric. Instead, we are here extracting from the Laplacian in a coordinate system its spectrum, which is the smaller set of invariants which are invariant under the full unitary group. It is for this reason that the spectrum of the Laplacian does not in general contain the full geometric information about the manifold. In effect, modding out by a group that is too large leaves one with a set of invariants that is too small. 

What strategies can be employed to obtain the full set of geometric invariants of a manifold? To mod out the diffeomorphism group directly is notoriously hard. To mod out the full unitary group, i.e., to collect the spectra of an operator such as the Laplacian, does provide us with some geometric invariants but generally not with a complete set of invariants. One possibility is to consider
not only the spectrum of the Laplacian on scalar fields, but also to consider the spectra of  differential wave operators for tensorial fields whose dynamics is rich enough to match that of the richness of the metric. Examples of such operators are Laplace-type operators on covariant symmetric 2-tensors (such as the metric), see \cite{Kempf_2010,kempf2018quantum}. This approach is not (yet) available here on causal sets\footnote{Causal sets are a special type of Lorentzian discrete geometry that will be introduced in Sec. \ref{sec:cstheory}.} because it is difficult to define higher rank tensors on causal sets. 

Instead, we are here considering two other alternatives: one strategy is to consider the spectra of multiple operators. In this case, each spectrum is generally an incomplete set of geometric invariants for the above discussed reasons: by calculating the spectrum of an operator we are modding out the full unitary group while we should have modded out only the smaller diffeomorphism group, i.e., here the permutation group. However, for each operator the set of geometric invariants that are missed can be different. Hence, the set of spectra of multiple operators may miss few, if any, geometric invariants. As we will show, the spectra of only few operators suffice to obtain a close to complete set of invariants. The other strategy that we pursue here is to consider small perturbations of the manifold, or rather, the causal set, by adding small attachments and to collect the spectra of the manifold and its perturbations. The idea here is again that the loss of geometric invariants when modding out the unitary group rather than the smaller diffeomorphism group is a different loss each time we do a small perturbation of the manifold. We demonstrate for causal sets up to the size that we were numerically able to study, that in this way a complete set of geometric invariants can indeed be acquired.

\subsection{Overview}
The spectra of operators such as the Laplace operator have long been considered in quantum gravity \cite{Benedetti:2009ge,Pinzul:2010ct,Kopf:1996vw}. For example, Gilkey and Hawking showed that in a path integral formulation of Euclidean quantum gravity, the action in terms of spectra reduces to the number of eigenvalues below the cutoff scale \cite{PhysRevD.18.1747, Gilkey}. Spectral geometric ideas have also entered in other models related to quantum aspects of Euclidean spacetime, such as non-commutative geometry, see e.g. \cite{connes,PhysRevLett.99.071302,PhysRevD.93.064034}. In this work, however, we consider Lorentzian spectral geometry. Compared to traditional studies, few results have been obtained on Lorentzian spectral geometry \cite{Kempf_2010} due to difficulties that arise as the elliptic Laplacian operator is replaced by the hyperbolic d'Alembertian.

In this work we propose and study new approaches for Lorentzian spectral geometry within the context of causal set theory. An earlier work \cite{SG1} by some of the present authors made  progress in this direction. There, the spectra of a number of operators constructed from the causal matrix (a type of adjacency matrix and a common lossless encoding of the information about a causal set) were considered, some of which (such as the d'Alembertian) are related to the theory of a scalar field on a causal set.  These were studied on the set of 6 and 7-element causal sets which one can fully enumerate. Since the causal matrix and d'Alembertian are non-normal and upper-triangular (with identical diagonal entries) matrices, they by themselves do not possess interesting spectra. In the absence of an analogue of the spectral theorem for non-normal operators, this leads to the fundamental question of how one can (spectrally or otherwise) extract the geometrically invariant information content of a non-normal operator. In \cite{SG1}, this question was studied by considering the spectra of the self-adjoint and anti-self-adjoint parts of the causal matrix, d'Alembertian and other non-normal operators. It was found that many of these operators indeed had good resolving power of the underlying causal sets. That work established spectral geometry in causal set theory as a fruitful avenue to pursue. However, many degeneracies still remained; any single operator produced several degenerate pairs of spectra belonging to physically distinct causal sets, calling for further study of geometric invariants in causal sets. 

In the current work, we build on these prior results and we suggest new ways to extract more geometric information from a causal set. We then test the new approaches on causal sets of up to 9 elements which implies a two order of magnitude increase in cardinality over prior work.

The paper is organized as follows: In Section \ref{sec:cstheory}, we introduce the reader to the basics of causal set theory and introduce the two strategies mentioned earlier. These strategies are then carried out in Sections \ref{sec:fixedset} and \ref{sec:pertset} which contain the main results of this work. In Section \ref{sec:outlook} we summarize our findings and discuss future directions.

\section{Causal Set Theory}\label{sec:cstheory}
\subsection{Causal Sets}
Causal set theory \cite{Bombelli:1987aa, SuryaRev} is an approach to quantum gravity that takes as fundamental the causal and Lorentzian spacetime structure of classical general relativity. Kinematically, a causal set consists of a set of discrete spacetime elements or spacetime ``atoms" and the binary (related or not) causal relations among them. Mathematically, a causal set is a locally finite partially ordered set. The causal relations among the elements is enough to reconstruct the metric up to a conformal factor. The volume of a region of spacetime is given by the number of elements in that region and this volume information gives the  metric's conformal factor \cite{osti_4057748, 1977JMP....18.1399M}. Hence the causal set tells us all we need to know about the classical spacetime and we no longer need a metric.

The full quantum dynamics of causal sets are envisioned to involve the (double) path integral or sum-over-histories approach \cite{Dowker_2010} and its development is a challenging work in progress. Meanwhile, if one ignores the backreaction between quantum fields and the causal set, one can study quantum fields living on a background causal set \cite{Johnston:2010su}. The spatio-temporal discreteness of the causal set covariantly cures the UV divergences of quantum fields \cite{Sorkin:2016pbz, Belenchia:2017cex}. In this paper, when we refer to quantum scalar field properties, we mean them within this context of a quantum field on a fixed background causal set. We will also however later comment on the application of our work to the full quantum path integral dynamics. For a recent review of causal set theory, see \cite{SuryaRev}.

A useful way to represent a causal set is by its causal matrix $C$, which is a kind of adjacency matrix, defined as
\begin{equation}
C_{xy}=\begin{cases}
1 & \text{for $x\prec ~y$}\\
0 & \text{otherwise}
\end{cases}
\end{equation}

where $\prec$ is the causal order relation such that $x\prec ~y$ means $x$ is causally related to and precedes a distinct element $y$. We have used the notation that the $xy$ indices refer to the matrix element relating elements $x$ and $y$.
By choosing a labelling such that earlier elements come before later ones\footnote{This ordering is called \emph{natural labelling} and is not in general  unique.}, the causal matrix $C$ can always be put into upper triangular form.

For a massless scalar field in $1+1$ dimensions, the retarded Green function $G_{ret}$ is the solution to 
\be
\Box G_{ret}(x,y)=\frac{\delta^2(x-y)}{\sqrt{-g}}.
\ee
It has support on the future lightcone such that it is zero unless $x\prec y$. In the causal set, $G_{ret}$ is simply $1/2$ of the causal matrix. 

 We also use the $2$d causal set d'Alembertian in this paper. We use the original definition \cite{SorkinLocality} which has no extra non-locality scale:

\begin{equation}
\frac{\ell^2}{4}B_{xy}=\begin{cases}
-1/2, & \text{for $x=y$}\\
1, -2, 1, & \text{for $n(x,y)=0, 1, 2,$ respectively, for $x\neq y$}\\
0 & \text{otherwise}
\label{box}
\end{cases}
\end{equation}

where $\ell$ is the discreteness scale, and $n(x,y)$ is the cardinality of the order-interval $\langle x, y\rangle=\{z\in\mathcal{C}|x\prec z\prec y\}$, or the number of elements of the causal set $\mathcal{C}$ that are causally between $x$ and $y$. This d'Alembertian acting on a constant field produces a term that approximates the Ricci scalar curvature. This has led to the d'Alembertian being used to propose an action (the Benincasa-Dowker action \cite{BDaction}) for a causal set.

\subsection{Non-normal Matrices}\label{nnmatrices}
Most of the matrices we consider in this work, for example the causal matrix $C$ introduced earlier, are non-normal. Non-normality of these matrices means that they do not commute with their conjugate transpose

\be
\left[A,A^\dagger\right]\neq 0.
\ee

This is equivalent to the fact that their self-adjoint (SA) and anti-self-adjoint (ASA) parts do not commute and hence cannot be simultaneously diagonalized. For a normal matrix, the spectral theorem applies and therefore the spectrum contains all the  retrievable basis independent information. This is certainly not the case for non-normal matrices. Moreover, as stated in the introduction, most of the matrices we consider can be put in triangular form with identical entries on the diagonal and therefore identical eigenvalues, hence their spectra are uninformative. In the case of the  d'Alembertian these diagonal entries take values one-half, and for the causal matrix they are zero-valued. 

For such non-normal matrices we can still consider the eigenvalues of their SA and ASA parts as  was explored in \cite{SG1}. But, as mentioned, this is not enough to recover the full information in these matrices. Given that the SA and ASA parts are diagonal in two different bases, one can deduce that some of the missing information is contained in the unitary transformation relating the two bases. This suggests approaches in which one still works with the SA and ASA parts of non-normal operators but tries to go beyond looking at eigenvalues alone. 
Take the causal matrix $C$ for example: a proposal would be to consider the canonical procedure where we a) diagonalize the SA part of $C$ and express the ASA part in the basis of the eigenvectors of the SA part, or b) diagonalize the ASA part of $C$ and express the SA part in the basis of the eigenvectors of the ASA part. We then compare both the spectra of the diagonalized matrix and the elements of the non-diagonal matrix across all causal sets of a given size $N$. A challenge in using this canonical approach is that there is an ambiguity in the representation of the eigenvectors. If the eigenvectors are normalized, then there is still an ambiguity in unit norm complex factors they can have. If there are degenerate eigenvalues, and there often are, then there are additional ambiguities stemming from being able to represent the eigenvectors as linear combinations of those that pair with the degenerate eigenvalues. However, even without bypassing this difficulty, we have empirically found that such a scheme would not succeed in fully encoding the information we are after. The  ambiguities make it difficult to identify all possible degeneracies among different $C$'s under this scheme, but we are still able to identify some, and that is enough to rule out this proposal. Perhaps another approach along these lines will still prove to be successful in future work.

In this work we look elsewhere to find the missing data. Next we give an overview of the strategies we explore in the remainder of this paper. 

\subsection{Two Strategies}
As we alluded to above, our main goal is to improve on previous results presented in \cite{SG1}. There, the spectra of the SA and ASA parts of different operators on all causal sets of 6 and 7 elements were studied. The study revealed that that Lorentzian spectral geometry on causal sets is not only viable but also effective. Nevertheless, different causal sets corresponding to different discrete spacetimes turned out to have identical spectra and hence could not be distinguished in that fashion. In this work we explore two possible strategies to go beyond these previous results.

 Our first strategy is to broaden the set of geometric invariants we consider. For example, we consider the collection of spectra from multiple operators on the causal sets, rather than the spectrum of an individual operator. A noteworthy new operator that we consider in this work is the commutator (for example of the self-adjoint and anti-self-adjoint parts of $C$). As mentioned already, the commutator is deeply connected to the non-normality of the matrices. We also explore geometric invariants that are non-spectral, such as the number of vertices. The common thread in this strategy is that we keep our sample space of causal sets (the set of $N$-element causal sets where $N\in[3,9]$) fixed and consider various combinations of invariants.
    
 Our second strategy is to consider a number of geometric invariants on our original set of $N$-element causal sets as well as on auxiliary higher cardinality causal sets. These auxiliary higher cardinality causal sets are constructed in a systematic and covariant manner by \emph{perturbing} the original causal sets. For example, we perturb a causal set by linking a new vertex to its future or its past. Another transformation we consider is gluing a causal set to a  time reversed copy of itself, a procedure we refer to as sandwiching. The sandwiching is motivated by the discovery of certain symmetries relating some causal sets with degenerate spectra as explained in Section \ref{sandw}. We then look at the resolving power of different geometric invariants under these kinds of transformations.
    
In the next two sections we elaborate on these two strategies.

\section{Keeping Causal Set Cardinality Fixed and Exploring Geometric Invariants}\label{sec:fixedset}

The first strategy that we explored was to simultaneously consider a number of different geometric invariants, for instance the spectra of multiple operators. By doing this, we improve the resolving power compared to the single operator case. However, the information we can glean from the operators saturates as we increase the number we consider. In other words, while considering the spectra of two or three operators can result in an improvement over considering the spectrum of any single operator, further improvement cannot be achieved by considering say ten more operators (at least not within the wide variety of operators we considered).  This is because the degeneracies of many individual operators overlap with one another.\footnote{Maybe not surprisingly, since the operators we consider can typically be expressed as different functions of the causal matrix, hence their similar capability to distinguish among causal sets.}

Below we present some methods to extract more spectral information from non-normal operators. In particular, we will consider the spectrum of the commutator between the SA and ASA parts and then other quantities which are invariant under relabeling of a causal set which are not typically derived as spectra of operators defined on discrete geometries.

\subsection{Commutator}
We found that a particularly useful set of operators to consider in conjunction with one another are the SA part, ASA part, and the commutator (Com) between them, for a given matrix. The commutator naturally captures some information about the transformation relating the eigenbases of the SA and ASA parts (see Section \ref{nnmatrices}). Below we will focus on this triplet of spectra for the d'Alembertian $B$ and the causal matrix $C$, that we found to have the highest resolving power among the options we considered.
 
As mentioned before, in $1+1$ spacetime dimensions, $1/2$ times the causal matrix is also equivalent to the retarded Green function of a massless scalar field, i.e. $G_{ret}=\frac{1}{2}C$. Therefore the matrix $C$ can be regarded either as the $G_{ret}$ for a massless scalar field, or more generally just the causal matrix. Similarly
we could view $B$ as a more general matrix given by the rules in \eqref{box}, rather than as a d'Alembertian in $2d$. By doing so, we are not constrained by their physical interpretation in a fixed  dimension and we can use them to probe spectral geometry in any dimension.

\begin{table}[h]
\label{comtable}
\begin{centering}
\begin{tabular}{|l|l|l|}
\hline
 & B & C \\ \hline
3-orders: 5            & 4 (80\%)  & 4 (80\%) \\ \hline
4-orders: 16           & 14 (87.5\%)  & 14 (87.5\%) \\ \hline
5-orders: 63           & 53 (84.13\%) & 53 (84.13\%) \\ \hline
6-orders: 318          & 286 (89.94\%) & 286 (89.94\%) \\ \hline
7-orders: 2045         & 1920 (93.89\%) & 1919 (93.84\%) \\ \hline
8-orders: 16999        & 16502 (97.08\%) & 16492 (97.02\%) \\ \hline
9-orders: 183231       & 180092 (98.23\%) &  179933 (98.2\%) \\ \hline
\end{tabular}
\caption{\label{tabcom} The number of unique equivalence classes of spectra for 3- to 9-orders considering the SA part, ASA part and their Com for the d'Alembertian $B$ and causal matrix $C$.  In brackets we have also included the ratio of unique spectra to the total number of spectra (enumerated in the first column).}
\end{centering}
\end{table}

In Table \ref{tabcom} we summarize the results from considering the spectra of the SA part, the ASA part and their Com for $B$ and $C$ and causal sets of size $3$ to $9$. This combination does not fully resolve all causal sets of a given cardinality, but we found that it has a high resolving power that improves with increasing causal set cardinality.
It is interesting to see that $B$ and $C$ have very similar resolving power, with the former doing slightly better. The inverse of the SA part of $G_{ret}$ has been considered as an alternative definition of the d'Alembertian. It was abandoned in favor of definition \eqref{box} due to its less apparent frame-independence \cite{SorkinLocality}.  We also considered this triplet of spectra for other operators such as the Feynman propagator, and found that the results were weaker than those obtained with $B$ or $C$.

Note that the commutator between the SA and ASA parts is a normal matrix, hence beyond its spectra no extra information can be extracted from it. We also empirically verified that considering the spectrum of the anti-commutator  in addition does not yield any benefit. Overall, we are able to distinguish slightly more than $98\%$ of all causal sets up to 9 elements using this method. We will further discuss the nature of some of the leftover degeneracies at the beginning of Section \ref{sandw}.

\subsection{Non-spectral Geometric Invariants}
In addition to spectra, there are other quantities that are invariant under relabelling of the causal set elements.  These invariants typically carry interesting information about graphs and they can be extracted from the adjacency matrix (causal matrix). They can be considered in trying to resolve the remaining degeneracies. Examples of these invariants include: the total number of elements, the number of edges, the number of initial and final elements of each set, the number of ingoing and outgoing edges, the number of disconnected parts of a given graph, and so on. What we found is that in some simple cases these quantities can uniquely characterize the sets without the need for any spectrum, and in other cases they help in partially distinguishing the leftover degeneracies. For instance, the majority of the degeneracies left after considering the spectra derived from the SA and ASA parts of $B$ and $C$ are given by time reversal pairs. One quantity that can be used to alleviate this problem is the number of initial and final vertices. If we consider these numbers jointly with the spectra of the operators constructed from $B$ and $C$, then causal sets of up to $N=5$ vertices can be fully distinguished. The situation also improves for the remaining sets as can be seen from Table \ref{tabcominv}.

\begin{figure}[h]
\centering
\includegraphics[width=.23\linewidth]{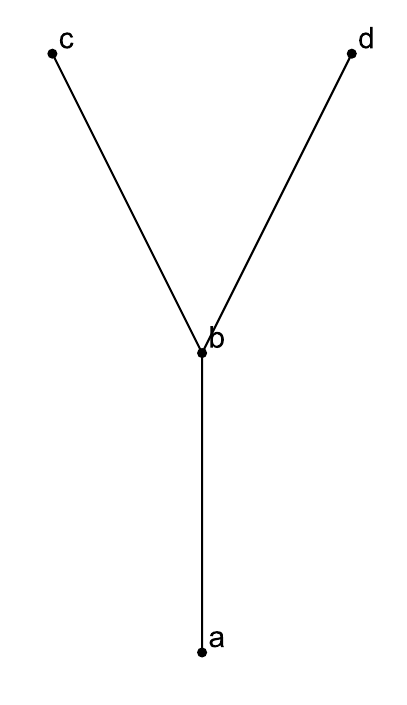}
\caption{Hasse diagram of a Y-shaped 4-element causal set. Time flows upwards and the lines show causal relations not implied by transitivity.}
\label{fig:ygraph}
\end{figure}

The quantity that we found to be most capable of resolving the remaining degeneracies after considering the spectra is the \textit{degree sequence}, defined as a list of pairs of integers describing the number of ingoing and outgoing edges for each vertex in a set \cite{diestel2006graph,trpevski2016graphlet}. For the causal set in Figure \ref{fig:ygraph}, the associated degree sequence is
\be
d_\textrm{seq}=(01,10,10,12).
\label{degseq}
\ee
The first number of each element in \eqref{degseq} is the number of ingoing edges while the second is the number of outgoing edges. Note that the list is sorted by magnitude. This quantity is enough to distinguish all the sets for $N=4$ and it seems to always perform better than all the other non-spectral quantities we considered combined. When combined with the spectra, all the causal sets up to $N=7$ can be distinguished, as shown in Table \ref{tabcominv}. Another quantity which is used in Table \ref{tabcominv} for the sake of comparison is the number of initial and final elements, that for the set in Figure \ref{fig:ygraph} would be $v_{in}=1$ and $v_{out}=2$. In general, the problem of finding a complete set of invariants that uniquely characterize graphs up to a certain number of vertices is an open question.\footnote{See the following references about the graph isomorphism problem \cite{doi:10.1063/5.0006891,toran2005,10.1145/321958.321963,babai2015graph,zemlyachenko1985}.}

\begin{table}[h]
\label{comtable2}
\begin{centering}
\begin{tabular}{|l|l|l|l|l|}
\hline
$N$ (\# of sets) & B, $v_{in}$, $v_{out}$ & C, $v_{in}$, $v_{out}$ & B, $d_\textrm{seq}$ & C, $d_\textrm{seq}$\\ \hline
3-orders: 5            & 5 (100\%)  & 5 (100\%) & 5 (100\%)  & 5 (100\%) \\ \hline
4-orders: 16           & 16 (100\%)  & 16 (100\%) & 16 (100\%)  & 16 (100\%) \\ \hline
5-orders: 63           & 63 (100\%) & 63 (100\%) & 63 (100\%)  & 63 (100\%) \\ \hline
6-orders: 318          & 314 (98.74\%) & 314 (98.74\%) & 318 (100\%) & 318 (100\%) \\ \hline
7-orders: 2045         & 2037 (99.61\%) & 2037 (99.61\%) & 2045 (100\%) & 2045 (100\%) \\ \hline
8-orders: 16999        & 16880 (99.3\%) & 16873 (99.24\%) & 16987 (99.93\%) & 16987 (99.93\%) \\ \hline
9-orders: 183231       & 182365 (99.53\%) & 182211 (99.44\%) & 183073 (99.91\%) &  183062 (99.91\%) \\ \hline
\end{tabular}
\caption{\label{tabcominv} The number of unique equivalence classes of spectral and non-spectral invariants for 3- to 9-orders. The spectra of the SA and ASA parts, and their Com for the d'Alembertian $B$ and causal matrix $C$ are considered along with the number of initial and final elements in the second and third columns and the degree sequence in the last two.  In brackets we have also included the ratio of unique equivalence classes to the total number of spectra.}
\end{centering}
\end{table}

\section{Modifying the Causal Set Cardinality  and Exploring Geometric Invariants}\label{sec:pertset}

\subsection{Sandwiching}\label{sandw}
\begin{figure}[h]
\begin{centering}
\includegraphics[width=.5\linewidth]{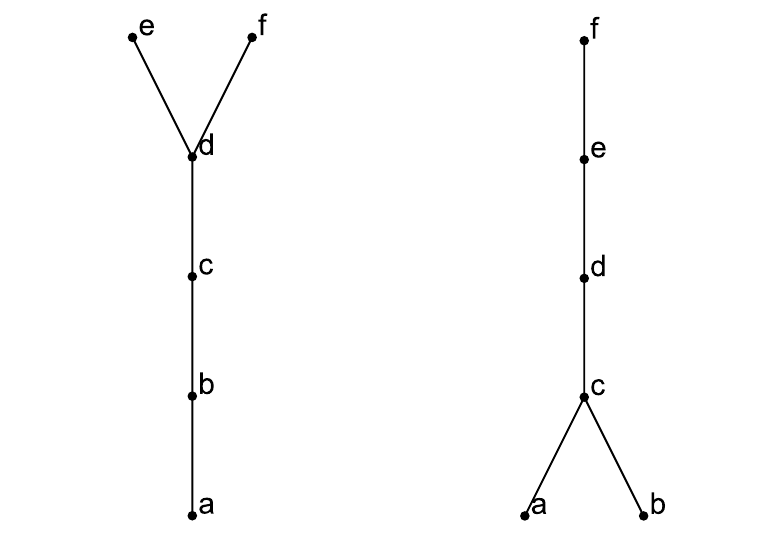}
\caption{\label{fig:timerev}A pair of 6-element causal sets related by time reversal.}
\end{centering}
\end{figure}

Upon inspection of the Hasse diagrams of causal set pairs typically yielding degenerate spectra for many different operators, we noticed that a large number of them were time reversals of one another. One such pair is shown in Figure \ref{fig:timerev}. Under time reversal, the causal matrix changes as $C\rightarrow C^T$, where $C^T$ is the transpose of $C$. Therefore it is evident that spectra from matrices such as the SA part of $C$ would not be able to distinguish such pairs because even at the matrix level they are invariant\footnote{We remind the reader that $C$ is real, therefore $C^\dagger=C^T$.} under the transformation $C\rightarrow C^T$. The ASA and Com parts of $C$ change sign under this transformation, so they would only distinguish such pairs if the eigenvalues did not come in $\pm$ pairs. This happens some of the time but not always. Hence, using these matrices we can only distinguish a subset of the time reversal pairs. More generally, even if the effect of $C\rightarrow C^T$ on the matrices is not as straightforward to see as in the cases just discussed, it seems to typically be the case that at the level of the spectrum we end up with some degeneracies involving time reversal related pairs of causal sets.

\begin{figure}[ht]
\begin{centering}
\includegraphics[width=.7\linewidth]{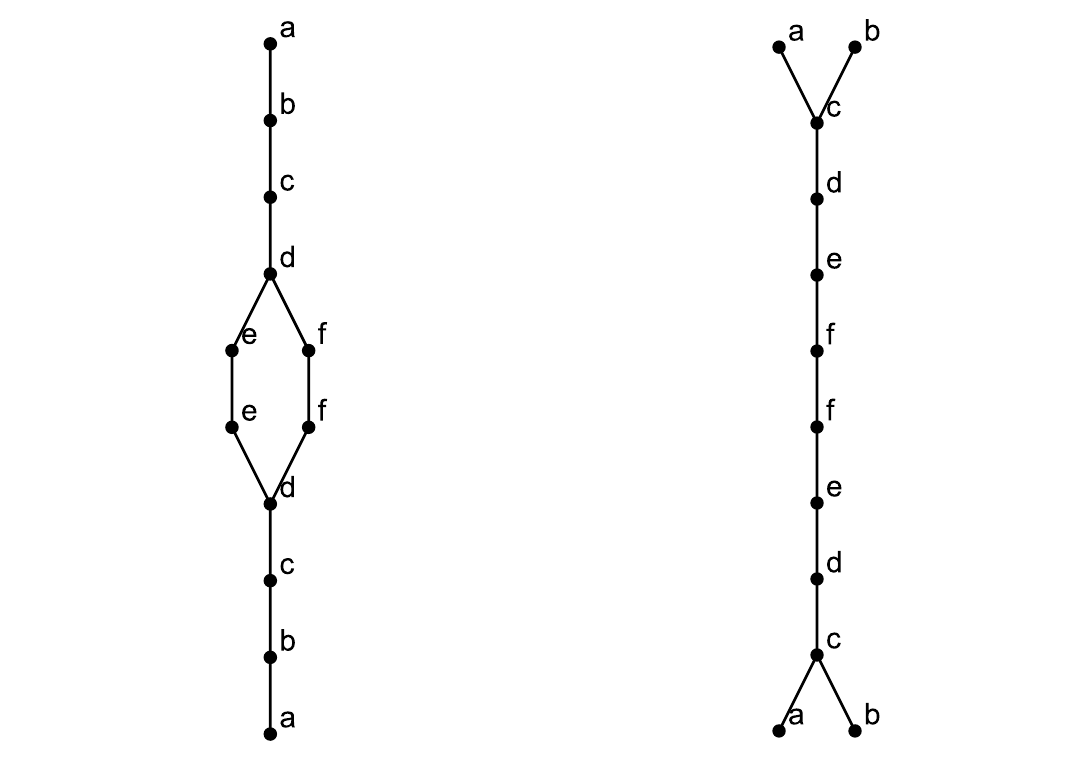}
\caption{\label{fig:sand}A pair of 12-element causal sets formed by sandwiching two 6-elements sets related by time reversal.}
\end{centering}
\end{figure}

This prompted us to consider the spectra of operators on composite causal sets formed such that the original time reversal relation between the pairs is broken. We form these composite causal sets via a ``sandwiching" process:
consider such a pair and call the two causal sets A and B. A has the same number of maximal elements (end points) as B's  minimal elements (initial points) and vice versa. This means that we can concatenate them. We do that by either gluing the minimal elements of B to the maximal elements of A or vice versa and adding an extra link where the gluing occurs. Doing this we  obtain two causal sets that we may call AB and BA. The additional link serves to make the sizes of the two composite causal sets be the same. If the original causal set had $N$ elements, then the sandwiched causal set will have $2N$ elements. Figure \ref{fig:sand} shows an example sandwich pair, where the original causal sets were those in Figure \ref{fig:timerev}.

To test how well we are breaking the degeneracies with sandwiching, we considered the spectra of combinations of 
operators derived from the causal matrix and its powers.\footnote{The entries of the matrix $C^n$ have a useful meaning: the value of each entry $C^n_{xy}$ gives the number of chains of length $n$ between element $x$ and $y$. Therefore in an $N$-element causal set, we would get all zero entries if we considered $C^n$ where $n\geq N$.} While we found no combination that possessed perfect resolving power, we found that already the combination of the spectra of $C+C^\dagger$ and $C^2+(C^2)^\dagger$, possesses very high resolving power with just two or three layer sandwiching. Table \ref{tab1} summarizes the number of unique (equivalence classes of) spectra from $C+C^\dagger$ and $C^2+(C^2)^\dagger$. The first row shows the number of unique spectra from considering the original causal sets only. The second row shows the number of unique spectra from considering the sandwiched causal sets as well. The third row shows the number of unique spectra when we additionally consider adding a third layer to the sandwiched causal sets.\footnote{In this three layer auxiliary causal set, we would end up with causal sets BAB and ABA where the same copy of a time reversal is glued to both the top and bottom of the other one in its time reversal pair.} In brackets we have also included the ratio of unique spectra to the total number of spectra (or causal sets) for each causal set cardinality.\footnote{The original cardinality.} The resolving power of the sandwiches and their 3-layer analogues slightly decreases with increasing causal set size.
\begin{table}[h]
\begin{centering}
\begin{tabular}{| l | c | c|c| r |}
 \hline
     \# of layers & 
    6-orders: 318 & 7-orders: 2045 & 8-orders:   16999 & 9-orders:  183231\\ \hline
    1 (original set) & 177 (55.66\%) & 1045 (51.10\%) & 8494 (49.97\%) & 89877 (49.05\%)\\  \hline
     2 (sandwich) & 318 (100\%) & 2044 (99.95\%) & 16977 (99.87\%) & $\sim$ 182835  (99.78\%)\\ \hline 
    3 (3-layers) & -- & 2045 (100\%)&  16989 (99.94\%) & $\sim$ 183078  (99.92\%)\\ \hline
\end{tabular}
   \caption{\label{tab1} The number of unique equivalence classes of spectra for 6- to 9-orders and their sandwiches. The first row shows the number of unique spectra from considering $C+C^\dagger$ and $C^2+(C^2)^\dagger$ in the original causal sets only. The second row shows the number of unique spectra from considering the sandwiched causal sets as well. The third row shows the number of unique spectra when we additionally consider adding a third layer to the sandwiched causal sets. In brackets we have also included the ratio of unique spectra to the total number of spectra.}
\end{centering}
   \end{table}

\subsection{Quill Perturbations}
A general way in which a causal set $\mathcal{C}$ can be modified in a basis independent manner is to systematically add extra elements and link them to pre-existing element(s). For example, one can add an element and link it to a maximal element, or to a minimal element, or to an intermediate element. In doing so, we are in a sense perturbing the original causal set. This procedure can be iterated to produce several auxiliary causal sets which can then equip us with additional spectra to use in spectral geometry. The sandwiching  in the previous subsection is a special case of this. 

\begin{figure}[h]
\begin{centering}
\includegraphics[width=.6\linewidth]{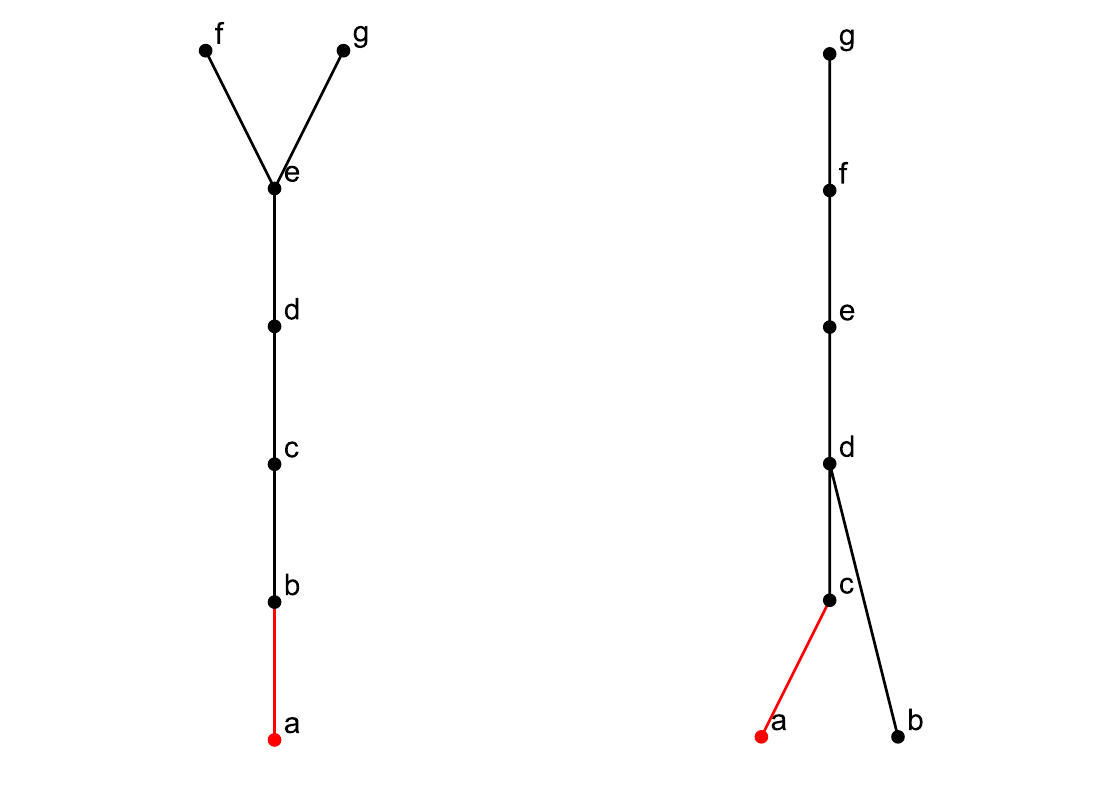}
\caption{\label{fig:quill}Quill-perturbing a pair of 6-element causal sets that were originally related by time reversal.  The extra element and link, highlighted in red, are added to a minimal element.}
\end{centering}
\end{figure}

 We call these perturbations quill perturbations because in the Hasse diagrams of the causal sets we are adding an edge. These edges look like the quills of a porcupine because they are attached at one end and stick out the other. Figure \ref{fig:quill} shows an example of a pair of causal sets (those that were in Figure \ref{fig:timerev}) that are quill perturbed by adding such perturbations to their minimal elements. The quills are highlighted in the figure. The unperturbed pair has degenerate spectra for operators such as $C+C^\dagger$ but the perturbed pair has non-degenerate spectra.

In Table \ref{perttab} below we summarize the results of a procedure using these perturbations that fully resolves all the $6$- to $9$-orders. The first row of the table below shows how many degenerate pairs remain after comparing the spectra of $C+C^\dagger$ and $C^2+(C^2)^\dagger$ as well as the number of maximal, minimal, and isolated elements. The second row shows the number of degenerate pairs that remain  after perturbing the remaining unresolved causal sets (from row 1) by adding a link to their maximal elements, and comparing the spectra of $C+C^\dagger$ and $C^2+(C^2)^\dagger$. The third row shows the number of degenerate pairs that remain after perturbing the remaining causal sets (from row 2) by adding a link to their minimal elements, and comparing the spectra of $C+C^\dagger$ and $C^2+(C^2)^\dagger$. The final row shows the number of degenerate pairs after perturbing the remaining unresolved causal sets (from row 3) by adding a link to one of the remaining elements that did not yet get a perturbation linked to them, and comparing the spectra of $C+C^\dagger$ and $C^2+(C^2)^\dagger$. We find that all causal sets of up to 9 elements can be uniquely distinguished using these quill perturbations.
   \begin{table}[h]
\begin{centering}
\begin{tabular}{| l | c | c|c| c |}
 \hline
    & 
    6-orders: 318 & 7-orders: 2045 & 8-orders:   16999& 9-orders:  183231\\ \hline
    No Perturbations & 33 & 283  & 2544  & 28733\\  \hline
      Perturb Maximal & 0 & 0& 4 &  270\\ \hline 
 Perturb Minimal & -- &  -- & 0 & 12\\\hline 
 Perturb All & -- &  -- & -- & 0\\ \hline
\end{tabular}
   \caption{\label{perttab} The number of degenerate spectra of $C+C^\dagger$ and $C^2+(C^2)^\dagger$ for 6- to 9-orders after considering   the number of minimal, maximal and isolated elements, as well as spectra from quill perturbations to each maximal, minimal and intermediate element.}
\end{centering}
   \end{table}
   
Finally, since we are dealing with discrete structures, let us discuss in what sense the term quill `perturbations' is justified. To this end, we now show that quill perturbations constitute not only small changes to the geometry of causal sets but also small changes to their spectra. 

We begin by noting that the quill perturbations can always be written as the addition of a first (or last) row and column to the unperturbed causal matrix (as well as its square,  SA and ASA parts). Thus, there exist a projection from the larger matrices onto the lower dimensional subspace of the unperturbed matrices.
This implies that 
the causal matrices before and after the perturbations, and also their squares, SA and ASA parts, satisfy the Cauchy interlace theorem \cite{Hwang}. This allows us to conclude that, for example, if the eigenvalues of $C+C^\dagger$ are $\lambda_1\leq\lambda_2,...,\lambda_N$ and the eigenvalues of the perturbed matrix are $\lambda'_1\leq\lambda'_2\leq...\leq\lambda'_{N+1}$, then $\forall j<N+1$, $\lambda'_j\leq\lambda_j\leq\lambda'_{j+1}$. 
The fact that for any causal set the eigenvalues are interlaced between the eigenvalues of any of its quill perturbations shows that, under any quill perturbation, the spectra remain close and are in this sense of interlacing `perturbations' of another.

The Cauchy interlace theorem bounds the intermediate eigenvalues of the perturbed matrix. Let us now show that there is also a bound to the change of the largest (in absolute value) eigenvalues. The 2-norm (spectral norm) of a matrix $A$ is given by the square root of the largest eigenvalue of $A^\dagger A$, i.e.,
\be
\norm{A}_2=(\sigma_{\textrm{max}}(A^\dagger A))^{1/2},
\ee
and, as a norm, it satisfies the triangle inequality
\be
\norm{A+B}\leq \norm{A}+\norm{B}.
\ee
Given a causal matrix $C$ of dimension $N$, we can add a new column and a new row of zeros, to make it a $N+1$-dimensional square matrix $\tilde{C}$. This is equivalent to adding a totally causally disconnected new element to the set.  We can then describe a quill-perturbed causal set by a matrix $\tilde{C}'=\tilde{C}+Q$, where $Q$ is a matrix with ones describing causal connections between the new element and the rest, and zeros everywhere else. One has analogous relationships for the square, SA and ASA parts of $\tilde{C}'$, giving for instance $\tilde{C}'_{SA}=\tilde{C}_{SA}+Q_{SA}$. The matrix $\tilde{C}_{SA}$ has an additional null eigenvalue compared to $C_{SA}$. One can then use the triangle inequality to bound the value of the largest eigenvalue of the perturbed matrix. The bound is
\be
\norm{\tilde{C}'_{SA}}_2\leq \norm{\tilde{C}_{SA}}_2+\norm{\phantom{\tilde{I}}Q_{SA}}_2.
\ee
The matrix $Q_{SA}$ is always a symmetric matrix with at most $2 N$ entries with value $1$. It is easy to show that its spectrum is going to be the pair $\pm \sqrt{\sum_{i,j=1}^N ((Q_{SA})_{ij})^2/2}$. Hence, the bound becomes
\be
\norm{\tilde{C}'_{SA}}_2\leq \norm{\tilde{C}_{SA}}_2+\sqrt{N}.
\ee
We now know that the largest eigenvalue of the perturbed matrix cannot be arbitrarily large and can only exceed the largest eigenvalue of the original unperturbed matrix by at most $\sqrt{N}$.  This construction is also valid for the ASA part of the causal matrix.

Considering that the entries of the square of the causal matrix represent the number of chains of length two between each pair of elements, a similar bound exists for $C^2$. In this case $Q_{SA}$ can have at most $2(N-1)$ non-zero entries, but now these non-zero entries can take  integer values in $\left[1,N-1\right]$.

In summary, this means that quill perturbations are perturbations not only geometrically but also spectrally. The eigenvalues are changing only within their prior spacing, by Cauchy's interlacing theorem, while the changes of the eigenvalues of the largest modulus are bounded by the triangle inequality. 

\section{Conclusions and Outlook}\label{sec:outlook}
We investigated methods of Lorentzian spectral geometry with causal sets. One approach to spectral geometry that we considered was to characterize a causal set by collecting a suitably large set of types of geometric invariants of this causal set. We found that if this set of invariants included the spectrum of the commutator of the self-adjoint and anti self-adjoint parts of matrices such as the causal matrix or d'Alembertian, then a high degree of resolution -- but not full resolution -- can be achieved. 

Another, novel, approach that we considered was to characterize a causal set by collecting a limited set of geometric invariants, but collecting them not only for the causal set in question but also for all of its quill perturbations, formed by adding relations to the original causal sets in a canonical way. We found that this set of geometrically invariant data succeeded in uniquely identifying all causal sets as far as we were able to check computationally, i.e., at least up to size $9$. 

Using properties of quill deformations, it may be possible to explicitly reconstruct a given causal set from its spectra. To see this, let us recall that, in continuum spectral geometry, it has been shown that in certain circumstances, the effectively linear relationship between small perturbations of  metrics and the correspondingly small perturbations of the spectra can be inverted to deduce small geometric changes from small spectral changes. Iterating the calculation of small shape changes from small changes of the spectra can then allow one to reconstruct geometries from spectra \cite{Aasen_2013,panine2016towards}. An efficient method is, for example, gradient descent down the landscape whose height is the $l^2$-distance between the current manifold's spectrum and the spectrum of the manifold that is to be reconstructed. Analogously, in the present paper, we showed that the small perturbations of a causal set imply perturbations of their spectra that are small in the sense of the Cauchy interlacing theorem. It should be very interesting to explore, therefore, whether a discrete analogue of a gradient descent method could then allow one to reconstruct causal sets from their spectra. 

Within causal set theory, there are several implications of our results. As mentioned in the introduction, each quantum gravity approach faces the task of characterizing gravitational degrees of freedom in a diffeomorphism invariant, or here permutation invariant, manner. Given that the geometric invariants we worked with had high resolving power of the underlying causal sets, they represent a promising candidate for this characterization. In the path integral for causal sets, the quantities one would sum over could be the invariants we have considered in this paper, such as the spectra of the quill perturbations. One may speculate that, depending on the choice of action, the path integral over the quill perturbations could in certain circumstances correspond to an integral over a derivative. In this way, a discrete analogue of the fundamental theorem of calculus, or its multidimensional generalization, the Stokes' theorem, may apply, thereby possibly leading to topological terms. 

For another potential use of our results, let us reconsider that when naively summing over all metrics or over all causal matrices, one redundantly sums also over metrics or causal matrices that describe the same spacetime up to a diffeomorphism, or permutation. Having a complete set of geometric invariants would  allow us to efficiently determine when two metrics or here two causal matrices are equivalent. This could allow one to fix a gauge by ensuring that the path integral sums over only one representative per equivalence class. In practice, it may be easiest to use the fact that  possessing a complete set of geometric invariants allows one to calculate the volume or cardinality of each gauge equivalence class. This means that in a path integral over all causal matrices, one can then apply the corresponding normalizing weights so that, in effect, each gauge equivalence class only contributes one weight to the path integral.  

This also touches on the question of which physical quantities (manifolds,  causal sets, or  spectra) to use as the representative of an equivalence class for the path integral. Such questions are also related to  \emph{spacetime functionalism}. In the functionalist approach of scientific philosophy, an entity would be called a spacetime depending on the  functional role it has; its functional role is in turn its role in the physical laws \cite{knox, baker}. In the case of this discussion, the functional role would be to provide  amplitudes for histories in the path integral.

Our results may also lend some insight into the problem of distinguishing manifoldlike causal sets from non-manifoldlike ones, as well as into the question why, for example in a path integral, manifoldlike causal sets would be more prominent than non-manifoldlike ones. One may speculate that manifoldlike and non-manifoldlike causal sets behave in some characteristic way differently under quill perturbations. For example, if the spectra of non-manifoldlike causal sets tend to change more than those of manifoldlike ones then this could imply that, in the path integral, the non-manifoldlike causal sets experience more destructive interference and are, therefore, playing a lesser role in the dynamics. 

Finally, some of our methods in this paper may possess applications also in other spectral geometry settings. In particular, the strategy of enriching the set of geometric invariants assigned to a spacetime by collecting also the geometric invariants of perturbations of the spacetime could be useful in the spectral geometry of continuous manifolds as well. 
For example, conformal perturbations of the metric on a compact Riemannian manifold without boundary can be described by a scalar function (and in the case of 2-dimensional compact Riemannian manifolds, all perturbations of the metric are conformal). 
Each scalar function describing a perturbation of the metric can be expanded canonically, i.e., diffeomorphism invariantly, in the eigenbasis of the Laplacian. This means that each eigenfunction of the Laplacian can be used to determine in a diffeomorphism invariant way a type of perturbation of the manifold, entailing a corresponding perturbation of the spectrum. Similar to collecting the spectra of the quill perturbations, we can now collect the spectra of these conformal perturbations to enrich the set of geometric invariants. It should be very interesting to study to what extent this expanded set of geometric invariants is able to distinguish compact Riemannian manifolds.

$$$$
\bf Acknowledgements: \rm This research was supported in part by Perimeter Institute for Theoretical Physics. Research at Perimeter Institute is supported by the Government of Canada through the Department of Innovation, Science and Economic Development and by the Province of Ontario through the Ministry of Research and Innovation. ML acknowledges the Fondazione Angelo della Riccia and the Foundation BLANCEFLOR Boncompagni-Ludovisi, n\'{e}e Bildt for financial support. YY acknowledges financial support from Imperial College London through an Imperial College Research Fellowship grant, as well as support from the Avadh Bhatia Fellowship at the University of Alberta. AK acknowledges support through the Discovery program of the National Science and Engineering Research Council of Canada (NSERC), the Discovery Project Program of the Australian Research Council (ARC) and through two Google Faculty Research Awards. 

\section*{References}
\bibliography{references}{}
\bibliographystyle{ieeetr}

\end{document}